\documentclass{JHEP3}

\usepackage{graphicx}
\usepackage{verbatim}
\usepackage{epsf}
\usepackage{latexsym}
\usepackage{amsmath,amsfonts,amssymb,amsthm}
\def\bseq{\begin{subequation}}  
\def\eseq{\end{subequation}}
\def\bsea{\begin{subeqnarray}}  
\def\esea{\end{subeqnarray}}

\newcommand{\bbox}{\lower.2ex\hbox{$\Box$}}

\newcommand{\beq}{\begin{equation}}
\newcommand{\eeq}{\end{equation}}
\newcommand{\bea}{\begin{eqnarray}}
\newcommand{\eea}{\end{eqnarray}}
\newcommand{\ena}{\end{eqnarray}}

\newcommand {\non}{\nonumber}

\renewcommand{\]}{\right]}

\newcommand{\be}{\begin{equation}}
\newcommand{\ee}{\end{equation}}

\preprint{
{\small{\textsf{SPHT-T08-141
}}}}

\title{\begin{center} 
Stringy Instantons
as Strong Dynamics
\end{center}}
\author{Antonio Amariti$^{1,a}$, Luciano Girardello$^{1,b}$ and Alberto Mariotti$^{2,3,c}$ 
\\ ~
\\
$^1$Dipartimento di Fisica, Universit\`a di Milano Bicocca\\
and \\
INFN, Sezione di Milano-Bicocca,\\ 
piazza della Scienza 3, I-20126 Milano, Italy\\
\\
$^2$
Service de Physique Theorique, SPhT \\
Orme des Merisiers, CEA/Saclay \\
91191 Gif-sur-Yvette Cedex, FRANCE \\
\\
$^3$
LPTHE, Universites Paris VI, Jussieu\\
F-75252 Paris, FRANCE\\
 \\
 $^a$\email{antonio.amariti@mib.infn.it} \\
$^b$\email{luciano.girardello@mib.infn.it} \\
$^c$\email{alberto.mariotti@lpthe.jussieu.fr}
}
\abstract{
We study the relation between stringy instantons 
and strong
dynamics effects
in type $II$B toric quiver gauge theories.
By exploiting the involutive 
property
of Seiberg
duality we relate the classical constraint on the moduli space of
the gauge theory 
with the
stringy instanton contribution to the superpotential.
The result holds for unitary, orthogonal and symplectic gauge groups.
}

\begin{document}

\section{Introduction}

Instantons are responsible for non-perturbative phenomena in $4D$
gauge theory and have, by now, found meaningful roles in string theory
as well.  Relevant work \cite{Dorey:2002ik} has been done in
different branches.  It has been shown that instantons in string
theory generate non perturbative contributions to the
superpotential \cite{storici}-\cite{Krefl:2008gs} and 
higher F-term contributions \cite{multifterm}.  Moreover
they have a role in model building, since they can be responsible
for moduli stabilization
\cite{modulisti} 
and for other phenomenological aspects like
neutrino masses,
supersymmetry breaking and gauge mediation \cite{feno}.  Other
results are found by adding fluxes \cite{fluxes} and more
generally by looking at the string compactification scenarios
\cite{altro}.

One distinguishes between ordinary $D$-brane instantons
and stringy $D$-brane instantons. 
Ordinary $D$-brane instantons are euclidean $D$-branes 
wrapping cycles in the geometry occupied
by other space time filling $D$-branes. 
They reproduce ordinary instantons effects for the gauge 
theory living on the space time filling $D$-branes.
Stringy $D$-brane instantons
are euclidean branes wrapped over cycles in the geometry
which are not occupied by any space-time filling 
brane.

In this note we investigate the relations between stringy instantons
and strong dynamics effects in type $IIB$ toric quiver gauge theories.
Stringy instanton contributions to the superpotential in quiver gauge
theories have been shown to exist for $SP(0)$, $SU(1)$ and $SO(3)$
nodes.  The second and third cases are considered stringy since the
low energy dynamics associated to those groups is trivial.  The
results, up to now, show that the stringy instanton contributions
reproduce the non perturbative part of the superpotential of the gauge
theory, i.e. part of the classical constraint on the moduli space.  Here we
present a clear-cut argument based on the involutive nature of the
Seiberg duality, which explains the retrieval of the exact
instanton contribution as a strong dynamical effect.  We shall speak
of equivalence or correspondence between the instantonic and gauge
schemes.

The paper is organized as follows.  First we give a general
overview of the correspondence between stringy instantons and
dynamical effects.  Then, in section 2 we review the one-instanton action
for a general quiver gauge theory and compute the contribution to the
superpotential.  In section 3 we argue that the correspondence is
implied by the involutive property of Seiberg duality.  In section 4
we give two examples, the $L^{121}$ 
and the $dP_1$ quiver gauge theories.  In section 5 we discuss the
correspondence for stringy instantons on $SP(0)$ and $SO(3)$ gauge
groups in orientifolded quiver gauge theories, with clarifying
examples.  Finally we conclude in section 6.  In appendix A we give
some details to complete the analysis of section 2.  In appendix B we
compute the bosonic integration over the zero modes.  In appendix C we
review some known result in our interpretation.

\subsection{Overview}
\label{Overview}

Consider a quiver gauge theory with an $SU(1)$ node and a tree level
superpotential $W_{tree}$.  
A stringy
instanton on a $SU(1)$ node gives rise to a superpotential term
\cite{GarciaEtxebarria:2007zv,Petersson:2007sc,Kachru:2008wt}. 
The cycle wrapped by the euclidean $D$-brane is occupied
also by one $D$-brane and the non trivial interaction lifts the
fermionic zero modes.
The resulting superpotential is
\be
\label{antoculo1}
W=W_{tree}+W_{inst}
\ee
Gauge theories with a $SU(1)$ gauge group are obtained 
as low energy (magnetic) 
descriptions of a strongly coupled $SU(N_c)$ gauge
theory with $N_c+1$ flavours.  The low energy description of this
strongly coupled $SU(N_c)$ gauge theory is a 
\emph{limiting case} of Seiberg
duality.  Indeed, it can be described by a magnetic gauge group
$SU(1)$, where the elementary degrees of freedom are mesons and
baryons. The baryons are the dual magnetic quarks.  The classical
moduli space of such a theory is not modified at quantum level.  The
classical constraint is imposed in the dual description by the
addition of a non trivial superpotential for the mesons and the
baryons, of the form 
\be
\label{antoculo2}
W_{eff} \sim B \mathcal{M} \tilde B - \det \mathcal{M}
\ee 
We shall show that the second term in (\ref{antoculo2})
is exactly reproduced by the stringy instanton contribution in 
(\ref{antoculo1}). 
Here and in the rest of the paper we set to unity 
the dimension-full coefficients.

Relations between non perturbative dynamics and
stringy instantons
has been already
observed in
\cite{Aharony:2007pr,GarciaEtxebarria:2007zv} 
for cascading gauge theories.
The correspondence we ascertain holds
at every step of a cascade when 
the Seiberg duality is 
in the limiting case.
The non perturbative contribution to the 
superpotential is then 
continuously mapped at every step
until the bottom of the cascade \cite{GarciaEtxebarria:2007zv}.

\section{Stringy instanton contribution}\label{generalact}

For the convenience of the reader we briefly review the basic instanton 
framework of relevance here.
We describe the most general configuration with
a $SU(1)$ node in a toric quiver gauge theory
and we place a stringy instanton on that node.
We consider only rigid instantons, 
without
adjoint fields charged 
under the $SU(1)$ gauge
group.
The instantonic action for toric quiver gauge theories
has been derived in 
\cite{Argurio:2008jm}. 
Toric quiver gauge theories can be obtained by performing
orbifold projection and higgsing
of the
$\mathcal{N}=4$ theory.
Along the same lines the instantonic action for toric quivers
can be derived from the
ADHM construction for 
$\mathcal{N}=4$ theory.

The system consists of $N$ $D3$ branes
and $k$ $D(-1)$ brane in type $II$ B.
We refer to \cite{Dorey:2002ik,revieu} for reviews.
The strings with endpoints attached to the $D3$ branes
lead to $SU(N)$ $\mathcal{N}=4$ SYM.
The strings  with endpoints attached to the $D(-1)$ branes 
lead to the neutral sector, uncharged under the gauge group.
It includes bosonic moduli $a^{\mu}$ and fermionic
zero modes $M^{\alpha \mathcal{A}}$ and $\lambda_{\dot \alpha \mathcal{A}}$
where $\alpha$ and $\dot \alpha$ denote the positive and
negative chirality in four dimension and $\mathcal{A}$ is an $SU(4)$
index (fundamental or anti fundamental) denoting the
chirality in the transverse six dimensions.
The equations of motion for the 
zero modes $\lambda_{\dot \alpha \mathcal{A}}$
implement the fermionic ADHM constraint.
There is also a triplet of auxiliary bosonic fields $D^c$ whose 
equations of motion implement the bosonic ADHM constraint.  
The charged sector is associated with strings stretching between
$D3$ branes and the $D(-1)$ branes.
It includes bosonic spinors $\omega_{\dot \alpha}$ and
$\bar \omega_{\dot \alpha}$ and fermions $\mu^{\mathcal{A}}$ 
and $\bar \mu^{\mathcal{A}}$.
These fields are matrices of dimension $N \times k$.

In order to obtain the toric quiver gauge theory 
together with the instanton sector 
the whole field content 
has to be
projected with the orbifold and then higgsed
in a consistent way 
\cite{Argurio:2008jm}.
Notice that instanton moduli 
scale with the same Chan-Paton structure of
ordinary gauge theory fields.

The resulting gauge theory is a toric quiver gauge theory
with many gauge groups, where
we can change the ranks of the groups by adding
fractional $D$-branes. The instanton sector works in a similar
way. There are $k$ instantons placed on each node,
and we can add instantonic fractional 
branes (not to be confused with fractional instantons) 
to obtain a different 
numbers $k_i$ of
instantons on the various nodes.  
Here we are interested in one instanton corrections without 
multi-instantons effects. 

\begin{figure}[h!!!]
\begin{center}
\includegraphics[width=7cm]{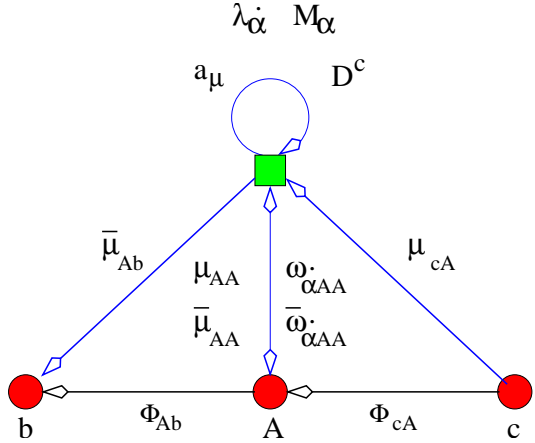}
\caption{Stringy instanton on a $SU(1)$ node in a generic quiver.
This is only the sector directly connected to the node $A$ of
an anomaly free quiver.}
\label{generale}
\end{center}
\end{figure}

From now on we consider
one instanton
placed on a $SU(1)$ node in a generic toric quiver gauge theory
(see Figure \ref{generale}).
We denote with $A$ the index associated with that node.
The auxiliary instanton group is $U(1)$. 
The node $A$ could be connected to the neighbor nodes with fields
$\Phi_{Ab}$, for outgoing arrows in the quiver, or with fields
$\Phi_{cA}$, for incoming arrows.  In general, there could be more
fields with the same gauge groups indexes.  To simplify the notation
we suppose here that every neighbor node is connected to the node $A$
with a single field.  The general case is treated in the appendix
\ref{species}.
\begin{table}[h!!!]
\begin{center}
\begin{tabular}{|c|c|c|c|}
\hline
Sector&ADHM&Statistic&Chan-Paton\\
\hline
Charged&$\mu_{Ab}$&Fermion&$k \times N_b$\\
\hline
Charged&$\bar \mu_{cA}$&Fermion&$N_c \times k$\\
\hline
Charged&$\mu_{AA}$&Fermion&$k \times N_A$\\
\hline
Charged&$\bar \mu_{AA}$&Fermion&$N_A \times k$\\
\hline
Charged&$\omega_{\dot \alpha \  AA}$&Boson&$k \times N_A$\\
\hline
Charged&$\bar \omega_{\dot \alpha \  AA}$&Boson&$N_A \times k$\\
\hline
Neutral&$a_\mu$&Boson&$k \times k$\\
\hline
Neutral&$M_{\alpha}$&Fermion&$k \times k$\\
\hline
Neutral&$\lambda_{\dot \alpha}$&Fermion&$k \times k$\\
\hline
Neutral&$D^c$&Boson&$k \times k$\\
\hline
\end{tabular}
\caption{Spectrum of the ADHM moduli in the charged and in the neutral sector}
\label{spettrot}
\end{center}
\end{table}

The spectrum is reported in Figure \ref{generale} and in Table 
\ref{spettrot}.
The toric quiver represents the gauge sector. The neutral sector
includes the bosonic zero mode $a_{\mu}$ and $D^c$, and the fermionic
zero modes $M^{\alpha}$ and $\lambda^{\dot \alpha}$ (only the $4$
component survive the orbifold projection in the one instanton case).
There is a charged sector 
connecting the node $A$ and the instanton, given by
$\omega_{\dot \alpha AA}, \bar \omega_{\dot \alpha AA}, \mu_{AA} ,\bar \mu_{AA}$,
and a charged sector connecting the instanton with
the neighbor nodes, in a way similar to the field
content of the gauge theory. 
For each existing outgoing arrow $\Phi_{Ab}$ there is a fermionic
zero mode
$\bar \mu_{Ab}$, and for each incoming arrow $\Phi_{cA}$ there is
$ \mu_{cA}$.

The instantonic action reads
\be
S_{inst}=S_1+S_2+S_W 
\ee
where
\be
S_1=i (\bar \mu_{AA} \omega_{\dot \alpha AA}+  
\bar \omega_{\dot \alpha AA} \mu_{AA} )
\lambda^{\dot \alpha}
- i D^c (  \bar \omega_{\dot \alpha AA}  
\tau^c   
\omega_{\dot \alpha AA}   )
\ee
\bea 
S_2
&=&
\frac{1}{2} \sum_{b} \big[ \bar \omega_{\dot \alpha AA}
\Phi_{Ab} (\Phi_{Ab})^{\dagger} \omega_{\dot \alpha AA}+
 i \, \bar \mu_{Ab} (\Phi_{Ab})^{\dagger} \mu_{AA} \big]+
 \\
 &&
\frac{1}{2} \sum_{c} \big[ 
\bar \omega_{\dot \alpha AA} (\Phi_{cA})^{\dagger}  
\Phi_{cA} \omega_{\dot \alpha AA}
-i \, \bar \mu_{AA} (\Phi_{cA})^{\dagger} \mu_{cA} \big] \eea
\be
\label{actiondersup}
S_W=
-\frac{i}{2}\sum_{b,c} \bar \mu_{Ab} 
\frac{\partial W}{\partial ( \Phi_{cA}  \Phi_{Ab})} \mu_{cA},
\ee
Observe that the $S_{W}$ action involves derivatives of the
superpotential with respect to bilinears of fields 
contracted on the $A$ 
index%
\footnote{This is necessary in order to take into account
 the contribution to this
 expression for 
 non abelian 
 superpotential
 and for
 superpotentials with terms 
 involving more than 3 fields.}.

The stringy instanton contribution is obtained by
integrating over all the zero modes
\be
Z=\mathcal{C} \int d\{ a_{\mu}, M, \lambda^{\dot \alpha},D,\omega_{AA} ,
\bar \omega_{AA} ,\mu_{AA},  \bar \mu_{AA},
\bar \mu_{Ab},  \mu_{c A} \} \, e^{-S_{inst}}
\ee
where $\mathcal{C}$ is a dimension-full parameter
which is discussed in appendix \ref{U1bosonic}.
The integration over the $a_{\mu}$ and the $M^{\alpha}$
zero mode is interpreted as the superspace integration.
Hence the stringy instanton 
contribution to the superpotential is given by
\be
W_{inst} \sim
 \int d\{ \lambda^{\dot \alpha}, D,\omega_{AA} ,
\bar \omega_{AA} ,\mu_{AA},  \bar \mu_{AA},
\bar \mu_{Ab},  \mu_{c A}  \} \, e^{-S_{inst}}
\ee
The bosonic integration is discussed in appendix \ref{U1bosonic}.
As for the fermionic integration
\be
W_{inst}
\sim
\int d \lambda^{\dot \alpha} d\bar \mu_{AA} \, d\mu_{AA} 
\prod_{b,c}  (d \bar \mu_{Ab})^{N_b} \, (d\mu_{cA})^{N_c} e^{-S_{inst}}
\ee
the integral on $\lambda^{\dot \alpha}$ 
can be performed as in \cite{Petersson:2007sc}
using the $S_1$ part of the instanton action
and it gives
the ADHM fermionic constraints. 
It also saturates the fermionic integration on
$\bar \mu_{AA}$ and $ \mu_{AA}$. 
We end up with the integral
\be
W_{inst}
\sim
\int  
\prod_{b,c}  (d \bar \mu_{Ab})^{N_b} \, (d\mu_{cA})^{N_c} e^{-S_{W}}
\ee
and this fermionic integration gives
\be
\label{genericinst}
W_{inst}
\sim
\det  \big(
\frac{\partial W}{\partial (\Phi_{cA} \Phi_{Ab})} \big)
\equiv
\det
\big(\mathcal{M} \big)
\ee
Notice that $\mathcal{M}$ is a 
a square matrix from the anomaly free condition
for the node $A$, which is
$
\sum_{b} N_{b}=\sum_{c}N_c
$.

\section{Discussion on the equivalence}

The contribution 
generated by a
stringy instanton on an $SU(1)$ gauge node 
is here obtained 
from the strong dynamics of the gauge theory.
This equivalence 
follows from the involutive property of the 
(limiting case) of Seiberg duality
(i.e. the case with $N_f=N_c+1$ for unitary gauge groups).

We consider the previous toric quiver gauge theory with a $SU(1)$ gauge group
labeled by $A$, with $N_f$ flavours spread on the nodes 
connected to the $SU(1)$ one.
The part of the superpotential involving
the fields charged under the gauge group $A$ 
is a generic holomorphic function
\be
\label{albec1}
W=W_0 (\Phi_{cA} \Phi_{Ab}, X^{(p)}_{bc} )
\ee
where the $\Phi$ fields are bifundamentals charged under the $SU(1)$.
The $X^{(p)}_{bc}$ are fields or products
of fields charged
under the gauge groups connected to $A$ in the quiver.
In section \ref{generalact} we showed that a stringy
instanton on node $A$ gives a
contribution to the superpotential of the form
\be
\label{albec2}
W_{inst}\sim\det \frac{\partial W_0 }{ \partial (\Phi_{cA} \Phi_{Ab})}
\ee

We now perform two consecutive Seiberg 
dualities on the node $A$
and compare the resulting theory with the original one.
The first step is a formal Seiberg
duality for the gauge group $SU(1)$.
This gives a $SU(\tilde N= N_f-1)$ gauge theory
with $N_f$ flavours, and superpotential
\be
W_{dual}=W_0 ( M_{cb}, X^{(p)}_{bc})+ M_{cb} q_{bA} q_{Ac}
\ee
where $M_{cb}=\Phi_{cA} \Phi_{Ab}$ and $  q_{bA}$ and $ q_{Ac} $
are the dual quarks.

The next step is another duality on the node $A$.  Since $N_f=\tilde
N+1$, the dual gauge group is $SU(1)$, and the superpotential is
\be
\label{albec3}
W_{eff}=W_0 ( M_{cb}, X^{(p)}_{bc})+ M_{cb} N_{bc} - N_{bc} b_{cA}
b_{Ab}+\det N_{bc} 
\ee 
where $N_{bc}= q_{bA} q_{Ac}$, the $b$ are
baryons, and we have changed the sign of the interaction term as in
\cite{Intriligator:1995au}.  The last two terms implement the
classical constraint on the moduli space \cite{Seiberg:1994pq}.

For the involution to hold,
this theory should coincide with the original one,
after integrating out the massive mesons $M_{cb}$, $N_{bc}$.
The equations of motions of the fields $N_{bc}$ give 
\be
b_{cA} b_{Ab}=M_{cb}=\Phi_{cA} \Phi_{Ab}
\ee
Hence we identify the baryons $b$ with the original fields $\Phi$.
The equation of motion of the meson $M_{cb}$ implies that
\be
\label{albec4}
N_{bc} \sim \frac{\partial W_0}{\partial M_{cb}}= \frac{\partial W_0}{\partial (\Phi_{cA} \Phi_{Ab}) }
\ee
so
we recover in (\ref{albec3}) the original theory (\ref{albec1})
and also the
stringy instanton contribution 
(\ref{albec2}),
i.e. the determinant term.
This proves the correspondence.
We conclude that the involution of the Seiberg duality
in the limiting case provides a gauge theory explanation
of the stringy instanton contribution.

\section{Examples}

In this section we exhibit two examples
of the correspondence:
a non chiral theory, the $L^{121}$
quiver gauge theory, 
and the $dP_1$ chiral theory.

We begin with a theory where there is a node with $N_f=N_c+1$,
we then consider strong dynamics for that node and
we study the low energy theory, 
performing a Seiberg duality in the limiting case,
obtaining
a non trivial contribution
to the superpotential.
The same contribution is obtained
analyzing directly the low energy theory and taking into account
the stringy instanton effect on the dualized node,
an $SU(1)$ node.

\subsection{Non chiral example: $L^{121}$}
\begin{figure}[h!!!]
\begin{center}
\includegraphics[width=5cm]{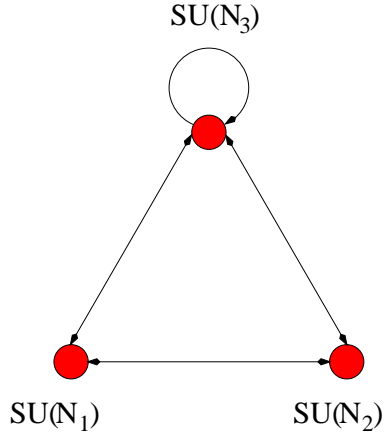}
\caption{$L^{121}$ quiver gauge theory.}
\label{SPPsemplice}
\end{center}
\end{figure}
\begin{figure}[h!!!]
\begin{center}
\includegraphics[width=5cm]{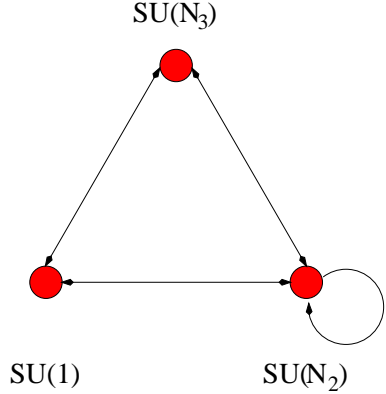}
\caption{$L^{121}$ after dualizing node 1.}
\label{SPPduale}
\end{center}
\end{figure}

The superpotential is
\be
W = - X_{33} Q_{31} Q_{13}  
  +  X_{33} Q_{32} Q_{23} 
  + Q_{21} Q_{13} Q_{31} Q_{12}
  -  Q_{32} Q_{21} Q_{12} Q_{23}
\ee
We choose 
the assignment
of ranks for the gauge groups such that
\be
N_2+N_3=N_1+1
\ee
We now consider strong dynamics for the node $1$.
This node has $N_f=N_c+1$. The low energy can be analyzed
performing a limiting case of Seiberg duality.
The magnetic gauge group is $SU(1)$, and the magnetic quarks
are identified with the baryons of the electric description.
The resulting theory has 
superpotential
\bea
W&=&-X_{33} M_{33}+X_{33} Q_{32}Q_{23}+M_{23}M_{32}-
M_{22}Q_{23}Q_{32}\nonumber \\
&&+M_{33}q_{31}q_{13}+M_{22}q_{21}q_{12}
-M_{23}q_{31}q_{12}-M_{32}q_{21}q_{13}
+\det{
\left(
\begin{array}{cc}
M_{22} & -M_{23}\\
-M_{32} & M_{33}
\end{array}
\right)
}
\eea
We have added the determinant contribution in order to
correctly implement
the classical constraint on the moduli space.
Integrating out the massive fields, we obtain the quiver in figure 
\ref{SPPduale} with the following superpotential
\be
\label{SPPdod}
W =  -q_{21} q_{13} q_{31} q_{12} + Q_{23} q_{31}  q_{13} Q_{32} 
  + q_{12} M_{22} q_{21} - Q_{32} M_{22}  Q_{23}
  +\det{
\left(
\begin{array}{cc}
M_{22} & -q_{21}q_{13}\\
-q_{31}q_{12} & Q_{32}Q_{23}
\end{array}
\right)
}
\ee
where there is an extra determinant term with respect
to the usual SPP superpotential. 
The theory of figure \ref{SPPduale}
has an $SU(1)$ gauge group. Strong dynamics effects
from the theory one step backward in Seiberg duality
have produced a non trivial superpotential contribution,
i.e. the determinant term in (\ref{SPPdod}).
We show here that the same term is  generated 
by a stringy instanton in the theory of figure \ref{SPPduale}.

\begin{figure}[h!!!]
\begin{center}
\includegraphics[width=8cm]{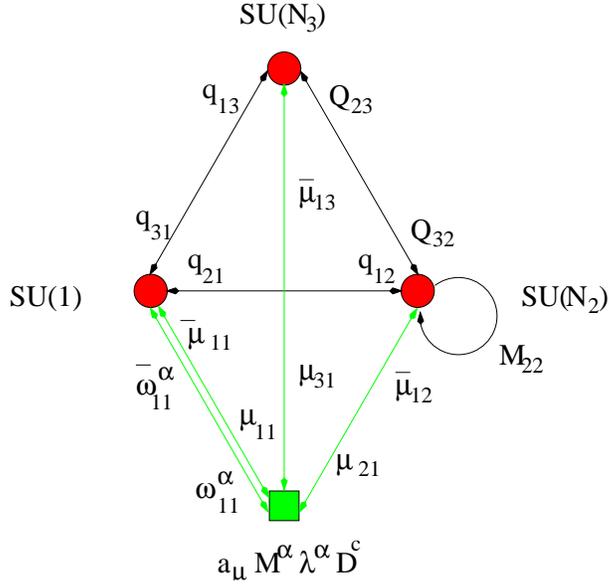}
\caption{$L^{121}$ with instanton on node $N_1=1$.}
\label{l131fig}
\end{center}
\end{figure}

The instantonic 
action for the $D$-instantons in the $SPP$ has been constructed in
\cite{Argurio:2008jm}:
\bea
 \label{azioneSPP}
S_{inst}= &&i (\bar \mu_{11} \omega_{\dot \alpha 11}+  
\bar \omega_{\dot \alpha 11} \mu_{11} )
\lambda^{\dot \alpha}
- i D^c (  \bar \omega_{\dot \alpha 11}  
\left(\tau^c\right)_{\dot \beta}^{\dot \alpha}   
\omega^{\dot \beta}_{11} ) 
\nonumber \\
&&
+
\frac{1}{2}  \bar \omega_{\dot \alpha 11}
\left( 
q_{12} q_{12}^{\dagger}
+
q_{21}^{\dagger} q_{21}
+
q_{13} q_{13}^{\dagger}
+
q_{13}^{\dagger}q_{31}
\right)
\omega_{\dot \alpha 11}
\nonumber\\
&&
+\frac{i}{2}
\left(
 \bar \mu_{12} q_{12}^{\dagger} \mu_{11} 
 +\bar \mu_{13} q_{13}^{\dagger} \mu_{11}
-\bar \mu_{11} q_{21}^{\dagger} \mu_{21}
-\bar \mu_{11} q_{31}^{\dagger}  \mu_{31}
\right)
\nonumber\\
&&
-\frac{i}{2}
\left(
\bar \mu_{12}    M_{22}     \mu_{21}
-\bar \mu_{12} q_{21}q_{13}  \mu_{31}
-\bar \mu_{13} q_{31}q_{12}  \mu_{21}
+\bar \mu_{13} Q_{32}Q_{23}  \mu_{31}
\right)
\eea
The corresponding quiver is given in
figure \ref{l131fig}.
The solid lines are the chiral superfields.
The dashed lines are the fermionic zero modes connecting 
the instanton and the other
$D$-branes in the theory. 
In order to compute the instanton contribution
to the
effective superpotential we have to integrate over
the fermionic and bosonic zero mode that couple 
among them and with the chiral superfields.
The $a_{\mu}$ and $M^{\alpha}$ zero modes
are the superspace coordinates.
The integral over the $\lambda^{\dot \alpha}$ and 
$D^c$ can be done using the 
first line in (\ref{azioneSPP}), and they give 
the two fermionic and the three fermionic ADHM constraints.
The other bosonic integration has been shown in appendix \ref{U1bosonic}
to give only a constant. 
The fermionic ADHM constraints are
\bea
&&
\delta\left(\bar \mu_{11} \omega_{\dot 1 11}
+ \bar \omega_{\dot 1 11} \mu_{11} \right)
\delta\left(\bar \mu_{11} \omega_{\dot 2 11}
+ \bar \omega_{\dot 2 11} \mu_{11} \right)
=
\left(\bar \mu_{11} \omega_{\dot 1 11}
+ \bar \omega_{\dot 1 11} \mu_{11} \right)
\left(\bar \mu_{11} \omega_{\dot 2 11}
+ \bar \omega_{\dot 2 11} \mu_{11} \right)
=\nonumber \\
&&
=
\bar \mu_{11}
\left(
\omega_{\dot 1 11}
\bar \omega_{\dot 2 11}
-
\omega_{\dot 2 11}
\bar \omega_{\dot 1 11}
\right)
\mu_{11}
\eea
This term saturates also the integrations over the zero modes 
$\mu_{11}$ and $\bar \mu_{11}$.

We are left with the following fermionic integral  
\be 
\label{fermint}
W_{inst}
\sim
 \int d^{N_2} \bar \mu_{12} 
d^{N_2} \mu_{21} d^{N_3} \bar \mu_{13} d^{N_3} \mu_{31} e^{-S_{inst}}
\ee
The relevant part of the action for this integral is the last line in
(\ref{azioneSPP}). It can be rearranged as
\be
S_{inst}=
\dots
-\frac{i}{2}
\left(
\begin{array}{cc}
\bar \mu_{12}&
\bar \mu_{13}
\end{array}
\right)
\left(
\begin{array}{cc}
M_{22}&-q_{21}q_{13}\\
-q_{31}q_{12}&Q_{32}Q_{23}
\end{array}
\right)
\left(
\begin{array}{c}
\mu_{21}\\
\mu_{31}
\end{array}
\right)
\ee
and the fermionic integration
(\ref{fermint}) gives the contribution 
\be
\label{albec10}
W_{inst}
\sim
\det{
\left(
\begin{array}{cc}
M_{22}&-q_{21}q_{13}\\
-q_{31}q_{12}&Q_{32}Q_{23}
\end{array}
\right)}
\ee
This is exactly the same determinant contribution 
we have obtained in (\ref{SPPdod}).
The correspondence between the superpotential terms
holds.
Indeed, adding (\ref{albec10})
to the tree level superpotential
for the quiver in figure \ref{SPPduale},
we exactly recover (\ref{SPPdod}).

\subsection{$dP_1$}
\begin{figure}
\begin{center}
  \includegraphics [width=10cm]{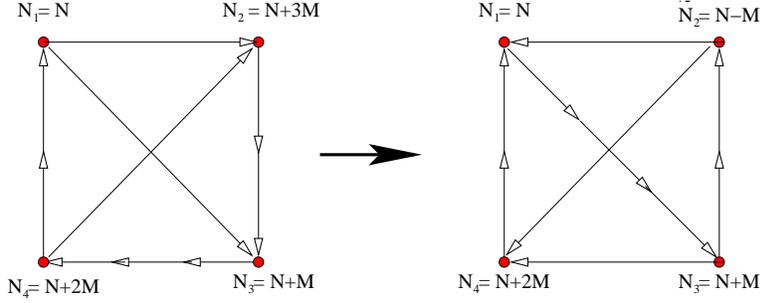}
\caption{Quivers representing the two dual phases studied for $dP_1$}
\label{dP1}
\end{center}
\end{figure}

Here we study the 
the chiral $dP_1$
toric quiver gauge theory.
The quiver of the theory is in figure \ref{dP1}.a .
The superpotential is
\beq
W = \epsilon_{\alpha \beta} X_{23}^{\alpha} X_{34}^{\beta} X_{42}
   +\epsilon_{\alpha \beta} X_{34}^{\alpha} X_{41}^{\beta} X_{13}
   -\epsilon_{\alpha \beta} X_{12} X_{23}^{\alpha} X_{34}^{3} X_{41}^{\beta}
\eeq
We choose the ranks to be 
$$
N_1=N \qquad N_2=N+3M \qquad 
N_3=N+M
\qquad
N_4 = N+2M
$$
and consider strong dynamics for the node 2.
The dual degrees of freedom
are the dual quarks ($b_{32}^{\alpha},b_{21},b_{24}$) and the mesons 
($M_{13}^{\alpha}, M_{43}^{\alpha}$).
The resulting quiver is in figure \ref{dP1}.b and
the superpotential, after integrating out the massive matter, is
\beq
\label{supdP1sec}
W = \epsilon_{\alpha \beta}  
b_{32}^{\alpha}b_{24} 
X_{41}^{\beta} X_{13} 
   +\epsilon_{\alpha \beta} M_{13}^{\alpha} b_{32}^{\beta} b_{21}
-\epsilon_{\alpha \beta} M_{13}^{\alpha} X_{34}^{3}X_{41}^{\beta}
\eeq
We used the equations of motion of the massive fields
($M_{43}^{\alpha}, X_{34}^{\alpha}$) that
fix
\be
M_{43}^{1}=X_{41}^1 X_{13}  
\qquad 
M_{43}^{2}=X_{41}^2 X_{13}
\ee
Choosing
$N=M+1$ we are in the limiting case of Seiberg duality,
where the dualized magnetic gauge group is $SU(1)$.
The classical constraint on the moduli space in this case 
is
implemented adding  to the superpotential (\ref{supdP1sec})
a determinant term
\be
\Delta W= 
\det \mathcal{M}=
\det
\left(
\begin{array}{cc}
M_{13}^{1}& -M_{13}^{2}\\
-M_{43}^{1}& M_{43}^{2}
\end{array}
\right)=
\det
\left(
\begin{array}{cc}
M_{13}^{1}& -M_{13}^{2}\\
-X_{41}^{1}X_{13}&X_{41}^{2}X_{13}
\end{array}
\right)
\ee
where we used the equation of motions to express it as a function of the
fields of the effective theory.

We now recover the same contribution 
as a stringy instanton effect in the magnetic theory,
the one in figure \ref{dP1}.b. 
We place a stringy instanton on the $SU(1)$ node. 
The saturation of the zero modes proceed as usual
and we are left with the following integral
\be
W_{inst} 
\sim
\int (d \bar \mu_{24})^{N_4} \,
(d \bar \mu_{21})^{N_1} \,
(d \bar \mu_{32}^{1})^{N_2} \,
(d \bar \mu_{32}^{2})^{N_2} \,
e^{-S_{inst}}
\ee
The relevant part of the instantonic action 
is (\ref{actiondersup}) and can be
deduced from the superpotential (\ref{supdP1sec})
to be
\be
S_{inst} 
\supset 
 \epsilon_{\alpha \beta} \bar\mu_{24} X_{41}^{\beta} X_{13} \mu_{32}^{\alpha}
+\epsilon_{\alpha \beta} \bar\mu_{21} M_{13}^{\alpha} \mu_{32}^{\beta}
=
\left(
\begin{array}{cc}
\bar \mu_{21}&
\bar \mu_{24}
\end{array}
\right)
\left(
\begin{array}{cc}
M_{13}^{1}&
-M_{13}^{2}\\
-X_{41}^{1} X_{13}&
X_{41}^{2} X_{13}
\end{array}
\right)
\left(
\begin{array}{c}
\mu_{32}^{2}\\
\mu_{32}^{1}
\end{array}
\right)
\ee
Performing the fermionic integrals we then find that
\be
W_{inst}
\sim
\det
\left(
\begin{array}{cc}
M_{13}^{1}& -M_{13}^{2}\\
-X_{41}^{1}X_{13}&X_{41}^{2}X_{13}
\end{array}
\right)
\ee
that is
$W_{inst}=\Delta W$
as claimed. The stringy instanton
contribution has been exactly 
mapped to the
strong dynamics effect.

\section{Orthogonal and symplectic gauge groups}
In this section we generalize the 
correspondence 
to orthogonal and symplectic gauge groups.
Quiver gauge theories with these groups can be obtained
from unitary gauge groups applying orientifold 
projections.
Since we consider toric quiver gauge theories, we use
the technology developed in \cite{Franco:2007ii}
to perform orientifold projections on dimer models \cite{dimerosi}.

The $O$-plane projects out some degrees of freedom
both in the gauge sector and in the instanton sector.
As a consequence the number of bosonic and fermionic
zero modes and the corresponding ADHM constraints 
are different \cite{Hollowood:1999ev}.

The stringy instanton contribution to the superpotential
for symplectic and
orthogonal gauge groups has been studied in 
\cite{Aharony:2007pr,GarciaEtxebarria:2007zv,Argurio:2007vqa,Krefl:2008gs}.
Non trivial contributions, in analogy
with the $SU(1)$ case, arise for stringy
instantons of $SP(0)$ and $SO(3)$ gauge groups.
The auxiliary instantonic groups are in these cases
$O(1)$ and $SP(2)$, 
respectively\footnote{In our convention $SP(2)\simeq SU(2)$.}.

The relation between stringy instantons
and strong dynamics effects of the gauge theory
holds also in these cases.
Electric magnetic dualities 
have been studied in 
\cite{Seiberg:1994pq,Intriligator:1995id,Intriligator:1995ne}
for symplectic and
orthogonal gauge groups.
For these groups there exist
limiting cases of the duality, as the
$N_f=N_c+1$ case for $SU(N_c)$.
They are respectively the
$N_f=N_c+4$ for $SP(N_c)$ and
$N_f=N_c-1$ for $SO(N_c)$ gauge groups.
For unitary groups the dual description
is a $SU(1)$ gauge theory. For symplectic and
orthogonal gauge groups
the dual descriptions are
$SP(0)$ and $SO(3)$, respectively.
Indeed they are the configuration
where stringy instanton effects add
to the superpotential.

We 
now 
point out the agreement
between stringy instanton 
and gauge theory analysis
with some example.

\subsection{The orthogonal case}

\begin{figure}
\begin{center}
\includegraphics[width=5cm]{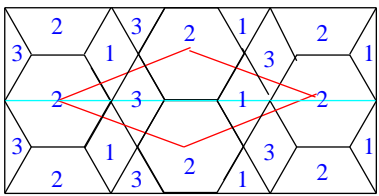}
\caption{Dimer model for the fixed line orientifold of the SPP.}
\label{SPPorientifold}
\end{center}
\end{figure}

\begin{figure}
\begin{center}
\includegraphics[width=5cm]{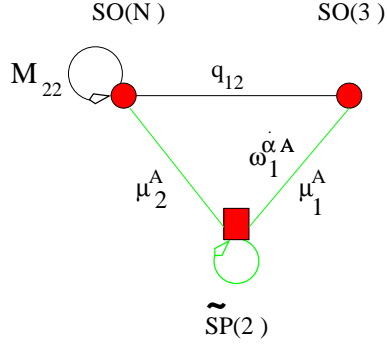}
\caption{Zero modes (in green) for the $\tilde{SP}(2)$
instanton placed on an $SO(3)$ gauge group. }
\label{figbell}
\end{center}
\end{figure}

In this subsection we study an orientifold projection 
of the SPP. We choose an orientifold from the
dimer with a fixed line, where the unit cell of the dimer
has a rhombus geometry (see Figure \ref{SPPorientifold}).
The orientifold charge for
the fixed line is chosen positive.
In this case all the unitary
groups $SU(N_i)$ become
orthogonal $SO(N_i)$ groups.
Half of the bifundamentals survive
the projection, and they become
\be
q_{i,j} = \left(\Box_i, \Box_j \right)
\ee
The adjoint field $M_{22}$ is projected 
to a symmetric representation.
We then choose the number of fractional branes
for each group such that
\be
N_1 = N \ \ \ \ 
N_2 = 3 \ \ \ \ 
N_3 = 0 
\ee 
This theory is described by the superpotential
\beq \label{diopor}
W = {q}_{12}^T {q}_{12} M_{22} 
\eeq
We add D-brane instantons
on the $SO(3)$ node.
In the ADHM construction 
of $SO(N)$ $\mathcal N = 4$ SYM
the instantonic auxiliary group
\cite{Hollowood:1999ev}
is $SP(k)$. In this case the counting of the zero modes
tells that stringy instanton  
contributes to the superpotential
if the auxiliary group is 
$SP(2)$.
The 
orientifolded quiver
gauge theory
with the instanton and the relative 
zero modes are shown in the figure \ref{figbell}.
The action for the zero modes is 
\bea \label{SO3action}
S_{inst} &=& 
i \lambda_{\dot \alpha}^i 
\left(
w_1^{\dot \alpha a} \sigma^{i}_{ab} \mu_1^b
\right)
+
i {\lambda'}_{\dot \alpha}^i \left(
\bar w_1^{\dot \alpha a} \sigma^{i}_{ab} \mu_1^b
\right)
-
i D^k_i
\left(
w_1^{\dot \alpha a}
\sigma^k_{\dot \alpha \dot \beta}
\sigma^i_{ab}
\bar w_1^{\beta b}
\right)
\nonumber\\
&+&\frac{1}{2}
\left(
w_1^{\dot \alpha a}
\left(
q_{12} q_{12}^{\dagger}
\right)
\bar w_1^{\dot \alpha a}
+i \mu_{2}^{a} q_{12}^\dagger \mu_{1}^{b} \epsilon_{ab}
-i \mu_2^{a} \epsilon_{ab} {\mu_2^T}^b M_{22}
\right)
\eea
where $a$ and $b$ are $SP(2)$ indexes.
Imposing the reality conditions we find six
independent  $\lambda_{\dot \alpha}^i ,{\lambda'}_{\dot \alpha}^{i}$
and four $M_{\alpha}$ zero modes. The $a_{\mu}$, in the adjoint of $SP(2)$,
are symplectic antisymmetric matrices. This representation
has dimension $1$,  which implies that there are four
zero modes from $a_\mu$. The $D_c$ are nine while there
are twelve independent $w_1^{\dot \alpha a}, \bar w_1^{\dot \alpha a} $ 
bosonic spinors. There are six fermionic $\mu_1^a$ fields 
connecting the gauge group $SO(3)$ with the
auxiliary instantonic group $SP(2)$. The sector connecting
the $SP(2)$ instanton with the flavor group gives $2N_f$
fermionic zero modes $\mu_2^a$.
We can now perform the integration over the fermionic and bosonic
zero modes to obtain the instanton contribution.
The $(a_\mu, M_{\alpha})$ zero modes are as usual interpreted
as superspace coordinates, giving the superpotential
contribution
\be
\label{SOsupinst}
W_{inst}=\mathcal{C} \int d\{\lambda, \lambda', D,\omega_{1} , \bar \omega_1, \mu_{1},
\mu_{2}  \} \, e^{-S_{inst}}
\ee
We discuss 
in the appendix \ref{SP2bosonic} the bosonic integration
and the dimension-full constant $\mathcal{C}$.
We only quote here the nine ADHM bosonic constraints
obtained integrating over the 
$D^k_i$
\bea \label{ADHMbos}
  \delta^{(9)}
\left(
w_c^{\dot \alpha a}
\sigma^k_{\dot \alpha \dot \beta}
\sigma^i_{ab}
\bar w_c^{\beta b}
\right)
\eea

Now we focus on the fermionic integration. 
The integration over the $\lambda_{\dot \alpha}^i ,{\lambda'}_{\dot \alpha}^{i}$
fermionic zero modes can be done using the
first two terms in (\ref{SO3action}) and it gives the six
ADHM fermionic constraints
\beq 
\label{ADHMferm}
  \delta^{(3)}(\omega_c^a \sigma^i_{ab} \mu_c^b)  
  \delta^{(3)}(\bar \omega_c^a \sigma^i_{ab} \mu_c^b)
\eeq
This saturate also the fermionic integration on $\mu_1^a$ in (\ref{SOsupinst}).
We are left with the fermionic integral
\be
 \label{intfermanto}
W_{inst}
\sim
\int \[d \mu_2^a \] e^{-S_{inst}}
\eeq
The integration is done expanding the relevant part of the
action in the exponent
\beq
S_{inst} 
\supset
 \mu_2^1 {\mu_2^T}^2 M_{22} - \mu_2^2 {\mu_2^T}^1 M_{22}
=\left(
\begin{array}{cc}
\mu_2^1& {\mu_2^T}^2\\
\end{array}
\right)
\left(
\begin{array}{cc}
0   & M_{22}\\
-M_{22}& 0
\end{array}
\right)
\left(
\begin{array}{c}
\mu_2^1\\
{\mu_2^T}^2
\end{array}
\right)
\eeq
The 
gaussian integration gives the contribution
\beq
\label{SP2supinst}
W_{inst} \sim \text{Pf}
\left(
\begin{array}{cc}
0   & M_{22}\\
-M_{22}& 0
\end{array}
\right)
 = 
\det M_{22} 
\eeq
This last equality holds since $M$ is a symmetric matrix.
In appendix
\ref{SP2bosonic}
 we show,
using dimensional analysis,  that the bosonic
integral is
adimensional. 
This means that it is independent from
the physical fields and it gives only
a constant contribution. 
We conclude that (\ref{SP2supinst}) is the
$SP(2)$ stringy instanton contribution
on the $SO(3)$ node. 

We now argue that the same relationship 
between stringy instanton and 
strong dynamics 
that holds in the case of
unitary groups is valid also in
this situation.

Once again we exploit the involutive property of Seiberg
duality.
We thus perform two consecutive Seiberg duality,
recovering in the end the starting theory. 
The first one
is a formal Seiberg duality on the $SO(3)$ node with $N$
flavours, and we obtain the theory one step backwards.
This gives
an $SO(\tilde N=N_f-N_c+4=N+1)$ gauge group with $N$ flavor.
The superpotential of this theory is 
\beq
W = Q_{12}^T Q_{12} N_{22} + N_{22} M_{22}
\eeq
Integrating out the massive field this superpotential
vanishes. 
We perform then another Seiberg duality.
Since for this theory $N_f=N_c-1$ we
are in the limiting case of Seiberg duality for orthogonal
gauge groups. The dual gauge group
is $SO(N_f-N_c+4=3)$ and the superpotential 
\be
\label{albecSO3}
W = q_{12}^T q_{12} M_{22} + \det{M_{22}}
\ee
where we have added the determinant to take into 
account the classical
constraint on the moduli space.

We have thus recovered in (\ref{albecSO3}) the starting superpotential 
(\ref{diopor}) and the stringy instanton
contribution (\ref{SP2supinst}).
So also for orthogonal gauge group
we have mapped the stringy
instanton contribution 
in strong dynamics effects.

\subsection{The symplectic case}

We 
first consider
symplectic SQCD.
We take $SP(0)$
as the gauge
group,
with $SP(N)$ flavours.
There is a meson $M$
in the antisymmetric representation
and there is no superpotential.

It has been shown that a 
stringy instanton 
on the $SP(0)$ gauge
group gives a non trivial
contribution to the superpotential.
In the ADHM construction 
the instantonic auxiliary group for a symplectic gauge group
is $O(k)$. 
The non perturbative contribution
is obtained if the instantonic number is $k=1$, 
with auxiliary group $O(1)$.
There are no fermionic and bosonic
ADHM constraint,  no $w$, $D$ and $\lambda$ fields. 
There are two $M_{\alpha}$ and
four $a_{\mu}$
which are interpreted as the superspace
coordinates.
The 
instantonic action is given by the interaction 
of the meson with the fermionic zero modes $\mu$ connecting the
$O(1)$ instanton and the flavor group
\be \label{SistPF}
S = -\frac{i}{2} \mu M \mu^T
\ee
The superpotential contribution is obtained
integrating over the $\mu$ fermionic
zero modes
\be
\label{SIcon}
W_{inst} \sim
\int d[\mu] e^{-S}=
 \text{Pf} M
\ee

Also for symplectic gauge groups we relate this
contribution to strong dynamics effects.
Through a formal electric magnetic 
duality
on
the $SP(0)$ node we obtain the dual theory.
It is an $SP(\tilde N=N_f-N_c-4=N-4)$ gauge group
with $N$ flavours and no mesons.
We then perform another duality obtaining
a $SP(N_f-N_c-4=N-\tilde N-4=0)$ gauge group,
where the only degree of freedom is the meson $M$. 
This is
the starting theory.
However,
since we are in the limiting case of Seiberg duality
for symplectic gauge group, we also obtain the 
following contribution to the superpotential
\be \label{WSP}
W_{eff} = \text{Pf} M
\ee
which implement the
classical constraints on the moduli space.
The equivalence between (\ref{SIcon}) and (\ref{WSP}) 
shows that, also 
for symplectic gauge groups,
the strong dynamic effect  
coincides with the stringy instanton
contribution to the superpotential.

\subsubsection{Example: Orientifold of the double conifold}
\begin{figure}
\begin{center}
\includegraphics[width=8cm]{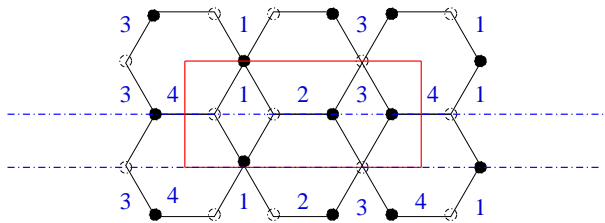}
\caption{Dimer for the orientifold of the double conifold. The dashed
  blue lines represents the orientifold fixed lines.}
\label{L222or}
\end{center}
\end{figure}
In this subsection we give an example of the correspondence using an
orientifold of the double conifolds.  The dimer model is represented
in figure \ref{L222or}. The unit cell is delimited by the red lines.
The projection is done by the two independent dashed blue fixed lines.
We choose their orientifold charge to be negative. This implies that
all the groups are symplectic. The bifundamentals fields are in the
$(\Box_i,\Box_j)$ representation of the $SP(2N_i) \times SP(2N_j)$
gauge groups. The fields in the adjoint representation of the
$SU(N_i)$ gauge groups are now in the antisymmetric representation of
the $SP(2N_i)$ groups.
The rank of the first group $SP(2N_1)$ is chosen to be zero.
The choice of the others ranks is free. Here, for simplicity, 
we choose the same rank $N$ for all of them.
The superpotential for this theory is 
\be 
\label{SPOTL222}
W = M_{22} \cdot Q_{23} \cdot Q_{23} 
  - Q_{23} \cdot Q_{23} \cdot Q_{34} \cdot Q_{34}
  + M_{44} \cdot Q_{34} \cdot Q_{34} 
\ee
where the $\cdot$ represent the symplectic products.

We add a stringy instanton on the
$SP(0)$ node
and we study its contribution to the superpotential.
\begin{figure}
\begin{center}
\includegraphics[width=5cm]{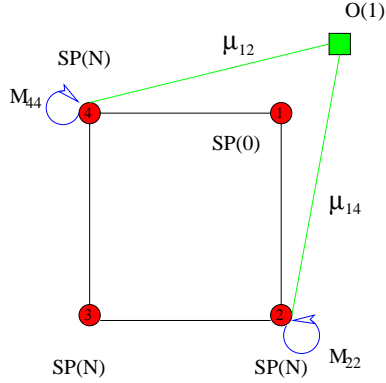}
\caption{ Stringy instanton on the orientifolded double conifold.
The green lines represent the fermionic zero modes in the 
instantonic action.}
\label{L22zeromodes}
\end{center}
\end{figure}
The zero modes are shown in figure \ref{L22zeromodes}.
The instantonic action 
is
\be
S_{\text{inst}} = -\frac{i}{2} \mu_{12}  M_{22} \mu_{12}^T - \frac{i}{2} \mu_{14}  
M_{44} \mu_{14}^T 
\ee
The integration over the fermionic zero modes $\mu_{12}$ and $\mu_{14}$ gives
a non perturbative contribution 
\be
\label{L222stinst}
W_{inst} 
\sim
\int d[\mu_{12}] d[\mu_{14}] e^{-S_{\text{inst}}}= \text{Pf}M_{22}\text{Pf}M_{44}
\ee
to the superpotential (\ref{SPOTL222}).

The same result can be found from the gauge theory
analysis.
The theory one step backwards in Seiberg duality 
is obtained 
by a formal duality on the $SP(0)$ node.
We get a
$SP(2N-4)$ gauge group with 
$2N$ flavours.
The superpotential of
 this dual theory is
\bea
W &=&   Q_{12} \cdot Q_{12} \cdot Q_{23} \cdot Q_{23}
  -   Q_{23} \cdot Q_{23} \cdot Q_{34} \cdot Q_{34}
\nonumber \\
  &+&   Q_{34} \cdot Q_{34} \cdot Q_{41} \cdot Q_{41}
  -   Q_{41} \cdot Q_{41} \cdot Q_{12} \cdot Q_{12}
\eea
We perform then another Seiberg duality to go back to the
starting theory.
We obtain a
$SP(0)$ gauge group and superpotential 
\be
W=-M_{24} \cdot M_{42}
  + M_{22} \cdot Q_{23} \cdot Q_{23} 
  - Q_{23} \cdot Q_{23} \cdot Q_{34} \cdot Q_{34}
  + M_{44} \cdot Q_{34} \cdot Q_{34} 
 + \text{Pf}
\left(
\begin{array}{cc}
M_{22}& M_{24}\\
M_{42}& M_{44}
\end{array}
\right) \ee 
The mesons are defined as $M_{ij}=q_{i \lambda_{i}}
J_{\lambda_{1} \lambda_{2}}q_{j,\lambda_{2}}$.  
Since we are in the case $N_f=N_c+4$
we have added the non perturbative term
to implement the classical constraint on the moduli space.
Note that the
antisymmetry of $J_{\lambda_1,\lambda_2}$ implies that $M_{42}= -
M_{24}^T$.  
Integrating out the massive fields 
$(M_{24},M_{42})$
we recover the starting theory (\ref{SPOTL222})
plus the 
non
perturbative contribution
\be
W_{np}=\text{Pf}M_{22}\text{Pf}M_{44}
\ee
This is exactly the
same contribution
obtained from the stringy instanton computation (\ref{L222stinst}).

\section{Conclusions}
In this paper we have 
considered stringy instantons in toric quiver gauge theories
deriving from $D3/D(-1)$ systems.
We have provided an interpretation
for the stringy instanton contribution as a strong dynamics effect
by analyzing the theory one step backwards in Seiberg 
duality, for the node where the instanton is located.
Our result is valid for stringy instantons on $SU(1)$, $SP(0)$
and $SO(3)$ 
nodes\footnote{See \cite{Intri2} for related 
discussion in the context of matrix models.}.

There are interesting aspects we have not discussed here.  The results
we presented could be extended to non toric quiver gauge theories.
Our analysis might also be useful in understanding the role played by
stringy instantons in dynamical supersymmetry breaking in quiver gauge
theories.  Another issue would be the study of non-rigid instantons
and multi-instantons effects in toric quiver gauge theories, and their
relation to strong dynamics of the gauge theory.  Finally\footnote{We
  thank A. M. Uranga for suggesting this to us.}  a similar
correspondence should exist for instantonic higher $F$-term
contributions in relation with strong dynamics leading to magnetic
$SU(0)$ gauge group, i.e. the $N_f=N_c$ case.

\section*{Acknowledgments}

We would like to thank 
R. Argurio,
G. Ferretti,
D. Forcella,
K. Intriligator,
S. McReynolds,
C. Petersson,
A. Zaffaroni
and especially
A. M. Uranga 
for stimulating comments and discussions.
A.~A.~ and L.~G.~ 
are supported in part by INFN, PRIN prot.2005024045-002 and the
European Commission RTN program MRTN-CT-2004-005104.
A.M. is supported by Fondazione Angelo Della Riccia.

\appendix

\section{General result for $U(1)$ instanton}\label{species}
In this appendix we compute the general contribution to the
superpotential for a rigid $U(1)$ instanton placed on a $SU(1)$ node
(denoted with $A$) in a toric quiver gauge theory,
generalizing the result of section \ref{generalact}.  The more general
configuration includes the possibility of having more than one field
with the same gauge group indexes, connected to the node $A$.  We
label these fields with an extra index $\alpha$ for outgoing arrow and
$\beta$ for incoming arrow.  Hence the fields connecting the node $A$
to the other gauge nodes are referred as $\Phi_{Ab}^{\alpha}$ or
$\Phi_{cA}^{\beta}$.  These extra indexes have to be inserted, and
summed over, in all the formula of section \ref{generalact}, e.g. for
the instantonic action.
An important remark is that now the anomaly free condition
for the node $A$ is 
\be
\label{anomalyfree}
\sum_{b,\alpha} N_{b}=\sum_{c,\beta}N_c
\ee
The procedure for getting the superpotential contribution is
as in section \ref{generalact}.
The integration of the bosonic zero modes and of the fermionic zero mode 
$\lambda^{\dot \alpha}$ and $\mu_{AA}, \bar \mu_{AA}, $ 
give the same result.
We have to perform the following integral
\be 
W_{inst}
\sim 
\int \prod_{b,\alpha,c,\beta} (d \bar
\mu_{Ab}^{\alpha})^{N_b} \, (d\mu_{cA}^{\beta})^{N_c} e^{-S_{W}} 
\ee
where now
\be
S_W=
-\frac{i}{2}\sum_{b,\alpha,c,\beta} \bar \mu_{Ab}^{\alpha} 
\frac{\partial W}{\partial ( \Phi_{cA}^{\beta}  \Phi_{Ab}^{\alpha})} \mu_{cA}^{\beta}
\ee
In order to compute this integral we can arrange the fermionic 
variable in vectors
\be
\bar \mu_{AB}= (\mu_{Ab}^{\alpha}) \qquad \mu_{CA}=(\mu_{cA}^{\beta})^{T}
\ee
of dimension
\be
B=1,\dots,\sum_{b,\alpha} N_b \qquad  
C=1,\dots,\sum_{c,\beta} N_c   
\ee
and rewrite the instantonic action as
\be
S_{W}=
-\frac{i}{2} \bar \mu_{AB} \mathcal{M}_{BC} \mu_{CA} 
\ee
where $\mathcal{M}$ is a matrix of dimension 
$
\sum_{b,\alpha} N_b \times \sum_{c,\beta} N_c   
$,
built taking derivatives of the superpotential
\be
\mathcal{M}=\big( \frac{\partial W}{\partial ( \Phi_{cA}^{\beta}  \Phi_{Ab}^{\alpha})} \big) 
\ee
$\mathcal{M}$ is a square matrix because of the anomaly free condition
(\ref{anomalyfree}).
The ordering of the fields in building $\mathcal{M}$ is irrelevant
for the final contribution to the superpotential, which is a determinant.
Indeed we can perform the fermionic integration and obtain
the stringy instanton contribution
\be
W_{inst}
\sim
\det  \big(
\frac{\partial W}{\partial (\Phi_{cA}^{\alpha} \Phi_{Ab}^{\beta})} \big)
\ee

\section{Bosonic integration}

In this appendix we show, via dimensional arguments
similar to \cite{Petersson:2007sc}, 
that the 
integration over the bosonic zero modes of the 
stringy instantons 
change the results of the fermionic integration
only by a constant factor.
We analyze the general bosonic integration for the $U(1)$ and
the $SP(2)$ instanton. The $O(1)$ case is trivial since there are no
bosonic zero modes to integrate over.

\subsection{The $U(1)$-instanton}
\label{U1bosonic}

In section \ref{generalact} we have considered an $SU(1)$ node $A$ in
the quiver and label with index $b$ all the outgoing arrows, and with
$c$ all the incoming arrows.  We have then the collection of fields
$\Phi_{Ab}$ and $\Phi_{cA}$.

We have seen that the contribution to the superpotential after
fermionic integration, due to an instanton on node $A$ is given by the
determinant of the squared matrix $\mathcal{M}$.  The determinant of
this matrix has mass dimension  
\be [\det \mathcal{M}]=M_s^{
  (\sum_{c} N_c) }=M_s^{ (\sum_{b} N_b) } 
\ee
We can now compute the dimension of the measure factor
for the general instanton computation of section \ref{generalact}
\be
\label{actiondim}
Z=\mathcal{C} \int d\{ a_{\mu}, M, \lambda,D,\omega_{AA} ,
\bar \omega_{AA} ,\mu_{AA},  \bar \mu_{AA},
\bar \mu_{Ab},  \mu_{c A}  \} \, e^{-S_{inst}}
\ee
The dimension-full coefficient $\mathcal{C}$ is for the moment unknown.
Using the usual standard dimensions
we arrive at
\be
[ d\{ a_{\mu}, M, \lambda,D,\omega_{AA} ,\bar 
\omega_{AA} ,\mu_{AA} , \bar \mu_{AA},
\bar \mu_{Ab},  \mu_{c A}  \}  ]=
M_s^{-n_a+\frac{1}{2}n_M-\frac{3}{2}n_{\lambda}+
2 n_D-n_{\omega,\bar \omega}+\frac{1}{2} n_{\mu ,\bar \mu}   }
\ee
Since
\be
n_a=4 \quad n_M=n_{\lambda}=2 \quad n_D=3 \quad n_{\omega ,\bar \omega}= 
4 N_A \quad n_{\mu \bar \mu}=2 N_A+\sum_{b} N_b+ \sum_{c} N_c
\ee
we obtain
\be
[ d\{ a_{\mu}, M, \lambda,D,\omega_{AA} ,
\bar \omega_{AA} ,\mu_{AA} , \bar \mu_{AA},
\bar \mu_{Ab}, \mu_{c A}  \}]
=M_s^{-(3 N_A-\frac{1}{2} ( \sum_{b} N_b +\sum_{c} N_c )   )}=M_{s}^{-\beta_A}
\ee
where we have recognized the 1 loop beta function of the node $A$.

Now, since $Z$ in (\ref{actiondim}) 
should be adimensional
we obtain that
\be
\mathcal{C}=\Lambda^{\beta_A}
\ee
Hence we have
\be
Z= \Lambda^{\beta_A} \int d\{ a_{\mu}, M, \lambda,D,\omega_{AA} ,
\bar \omega_{AA} ,\mu_{AA} , \bar \mu_{AA},
\bar \mu_{Ab}, \mu_{c A}  \} \, e^{-S_{inst}}
\ee
Now, we expect that
\be
Z=\int d^4 x d^2\theta \, W_{inst}
\ee
and then
\be
W_{inst}
=\Lambda^{\beta_A} \int d\{ \lambda,D,\omega_{AA} ,
\bar \omega_{AA} ,\mu_{AA}  ,\bar \mu_{AA},
\bar \mu_{Ab},  \mu_{c A}  \} \, e^{-S_{inst}}
\ee
Now, we have seen that the fermionic (plus the $D$) 
integrations give, when $N_A=1$, the following 
\be
W_{inst}=\Lambda^{\beta_A} I_{boson}\det \mathcal{M}
\ee
where $\mathcal{M}$ is the meson built before and we still have to perform
the bosonic integration $I_{boson}$,
and show that 
it gives a numerical coefficient.
Indeed the dimensional analysis gives
\bea
[W]&=&[\Lambda^{\beta_A}]+[I_{boson}]+[\det \mathcal{M}]
=\beta_A+[I_{boson}]+(\sum_{b} N_b)=\\
&=&
3-  \frac{1}{2} ( \sum_{b} N_b +\sum_{c} N_c ) 
+[I_{boson}]+(\sum_{b} N_b)=3+[I_{boson}]   \non
\eea
where we have used the anomaly free condition.
In order to have a superpotential of dimension 3 we have to
set $[I_{boson}]=0$, i.e. a number.

\subsection{The $SP(2)$-instanton}
\label{SP2bosonic}

We can easily repeat the analysis done in the previous section
for the $SP(2)$ instanton on the $SO(3)$ gauge node.
We denote with $A$ the $SO(3)$ gauge group
where we place the instantons
and label with index
$b$ all the outgoing arrows, and with $c$ all the
incoming arrows. 
In general the contribution to the
superpotential after fermionic integration,
due to $SP(2)$ instantons on node $A$
is given by a pfaffian 
of dimension
\be
[\text{Pf} \mathcal{M}]=M_s^{ (\sum_{c} N_c) }=M_s^{ (\sum_{b} N_b) }
\ee
We can now compute the dimension of the instanton measure factor
\be
\label{actiondimsp}
Z=\mathcal{C} \int d\{ a_{\mu}, M, \lambda,D,\omega_{AA} ,\mu_{AA},
\bar \mu_{Ab},  \mu_{c A}  \} \, e^{-S_{inst}}
\ee
The dimension-full coefficient $\mathcal{C}$ is up to now unknown.
Using the usual dimensions 
we arrive at
\be
[ d\{ a_{\mu}, M, \lambda,D,\omega_{AA} ,\mu_{AA} ,
\bar \mu_{Ab},  \mu_{c A}  \}  ]=
M_s^{-n_a+\frac{1}{2}n_M-\frac{3}{2}n_{\lambda}+
2 n_D-n_{\omega,\bar \omega}+\frac{1}{2} n_{\mu ,\bar \mu}   }
\ee
Now we have to remind that the auxiliary group for the instanton is
$SP(2)$ and this gives different numbers of components 
respect to the $U(1)$ case, that is
\be
n_a=4 \quad n_M=2, \quad n_{\lambda}=6 \quad n_D=9 \quad n_{\omega}= 
4 N_A \quad n_{\mu \bar \mu}=2 N_A+\sum_{b} N_b+ \sum_{c} N_c
\ee
we obtain
\be
[ d\{ a_{\mu}, M, \lambda,D,\omega_{AA} ,
\mu_{AA},
\bar \mu_{Ab}, \mu_{c A}  \}]
=M_s^{-(3 N_A-6-\frac{1}{2} ( \sum_{b} N_b +\sum_{c} N_c )   )}=M_{s}^{-\beta_A}
\ee
where we have recognized the 1 loop beta function of the $SO(N_A)$ node.

Now, since $Z$ in (\ref{actiondimsp}) 
should be adimensional
we obtain that
\be
\mathcal{C}=\Lambda^{\beta_A}
\ee
Hence we have
\be
Z= \Lambda^{\beta_A} \int d\{ a_{\mu}, M, \lambda,D,\omega_{AA} ,\mu_{AA} ,
\bar \mu_{Ab}, \mu_{c A}  \} \, e^{-S_{inst}}
\ee
Now, we expect that
\be
Z=\int d^4 x d^2\theta \, W_{inst}
\ee
and then
\be
W_{inst}=\Lambda^{\beta_A} \int d\{ \lambda,D,\omega_{AA}  ,\mu_{AA}  ,
\bar \mu_{Ab},  \mu_{c A}  \} \, e^{-S_{inst}}
\ee
Now, we have seen that the fermionic (plus the $D$) 
integrations give, when $N_A=3$, the following 
\be
W_{inst}=\Lambda^{\beta_A} I_{boson}\text{Pf} \mathcal{M}
\ee
where 
$I_{boson}$
is
the bosonic integration.
The dimensional analysis told us that
\bea
[W]&=&[\Lambda^{\beta_A}]+[I_{boson}]+[\text{Pf} \mathcal{M}]
=\beta_A+[I_{boson}]+(\sum_{b} N_b)=\\
&=&
3 N_A-6-  \frac{1}{2} ( \sum_{b} N_b +\sum_{c} N_c ) 
+[I_{boson}]+(\sum_{b} N_b)=3 N_A-6+[I_{boson}] \non
\eea
Since we have $N_A=3$,
in order to have a superpotential of dimension 3 we have to
set $[I_{boson}]=0$, i.e. a number.

\section{Relation with known models}

In this appendix we show that there
is no disagreement
between the stringy instanton 
contributions of 
\cite{Petersson:2007sc,Argurio:2007vqa}
and our results.

\subsection{The $SU(1)$ theory} 

The theory studied in \cite{Petersson:2007sc}
is the $\mathbb{C}^3/(\mathbb{Z}_2\times\mathbb{Z}_2)$
orbifold. This is a quiver gauge theory with four gauge groups,
described by the superpotential
\bea \label{starting}
W 
&&
=\Phi_{12} \Phi_{23} \Phi_{31}
-\Phi_{13} \Phi_{32} \Phi_{21}
+\Phi_{13} \Phi_{34} \Phi_{41}
-\Phi_{14} \Phi_{43} \Phi_{31}
\nonumber \\ 
&&
+\Phi_{23} \Phi_{34} \Phi_{42}
-\Phi_{24} \Phi_{43} \Phi_{32}
+\Phi_{12} \Phi_{24} \Phi_{41}
-\Phi_{14} \Phi_{42} \Phi_{21}
\eea
The ranks of the groups are $(N_1,N_2,N_3,N_4)=(N_1,N_2,1,0)$.
A stringy instanton placed on node 
$N_3$
contributes to the superpotential only if $N_1=N_2$.
In this case
it has be shown that its contribution is 
\be
\label{instpete}
W_{inst} = \det{\Phi_{12}} \det{\Phi_{21}}
\ee

We now find the same result 
from gauge theory analysis.
Dualizing the node $3$ we find 
a theory with gauge group
$SU(\tilde N_3=  N_1+N_2 -1)$ 
and superpotential
\be
W = M_{11} Q_{13} Q_{31} - M_{22} Q_{23} Q_{32}
\ee
We then dualize again node $3$.
Since it is in the case $N_f=N_c+1$,
 the dual theory has
$SU(N_3)=SU(1)$ gauge group, 
and the superpotential
is
\be
W =  M_{11} \Phi_{11} - M_{22} \Phi_{22}
   + \Phi_{11} \Phi_{13} \Phi_{31} - \Phi_{22} \Phi_{23} \Phi_{32}
   + \Phi_{12} \Phi_{23} \Phi_{31} - \Phi_{13} \Phi_{32} \Phi_{21}
   + \det
     \left(
     \begin{array} {cc}
     \Phi_{11}&\Phi_{12}\\
     \Phi_{21}&\Phi_{22} 
     \end{array}
     \right)
\ee
After the integration of the massive field $M_{11}$,$M_{22}$,$\Phi_{11}$,$\Phi_{22}$, 
the superpotential is 
\be
W =  \Phi_{12} \Phi_{23} \Phi_{31} - \Phi_{13} \Phi_{32} \Phi_{21}
   + \det
     \left(
     \begin{array} {cc}
     0&\Phi_{12}\\
     \Phi_{21}&0 
     \end{array}
     \right)
\ee
The first two terms are the same than (\ref{starting}).
The $\det$ piece coincide with (\ref{instpete}), 
as expected.
Note that it vanishes if $N_1 \neq N_2$ as in the stringy instanton
computation.

\subsection{The $SP(0)$ theory}

\begin{figure}[h!!!]
\begin{center}
\includegraphics[width=5cm]{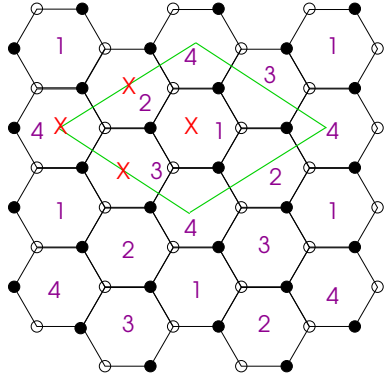}
\caption{Unit cell and fixed point for 
$\mathbb{C}^3/(\mathbb{Z}_2\times\mathbb{Z}_2)$}
\label{orientiargurio}
\end{center}
\end{figure}

It is also possible to make an orientifold projection of
the  $\mathbb{C}^3/(\mathbb{Z}_2\times\mathbb{Z}_2)$
orbifold.
A possible orientifold is described by the dimer in
Figure \ref{orientiargurio}.
It is a fixed point projection. Since $N[W]=8$, the total orientifold
charge is positive. This condition can be imposed choosing 
all the single charge to be negative. All the groups are identified with
themselves and they are all symplectic.
All the fields are bifundamental in the $(\Box_i,\Box_j)$ of
the $SP(N_i)\times SP(N_j)$ gauge groups. 
There is no superpotential for these fields.
This is the same projection described in \cite{Argurio:2007vqa}.
In that paper the ranks were $(N_1,N_2,N_3,N_4)=(N,N,0,0)$, 
and the stringy instanton was located on the third node.
The stringy instanton
 contribution to the superpotential is given by 
\be \label{ARGSP}
W_{inst}=\det{\Phi_{12}}
\ee
The same result can be found by the gauge theory analysis.
The dual theory 
has rank $2N-4$ for the third node, and superpotential
\be
W = M_{11} \cdot Q_{13} \cdot Q_{13} -  M_{22} \cdot Q_{23} \cdot Q_{32}
\ee
We then perform again electric magnetic
duality on the third node.
The gauge group
becomes $SP(N_1+N_2-N_3-4)=SP(0)$, and the superpotential
is
\be
W = M_{11} \cdot \Phi_{11} -  M_{22} \cdot \Phi_{22} + 
\text{Pf}\left(
\begin{array} {cc}
\Phi_{11} & \Phi_{12}\\
\Phi_{21} &\Phi_{22}
\end{array}
\right)
\ee
where all the blocks of the meson are in
an antisymmetric representation of the
flavor, and $\Phi_{21}=-\Phi_{12}^T$.
Integrating out the massive fields the
only
non vanishing term of the superpotential is
the non perturbative one
\be
W = \text{Pf} \left(
\begin{array} {cc}
0 & \Phi_{12}\\
-\Phi_{12}^T &0
\end{array}
\right) = \det{\Phi_{12}}
\ee
It is exactly 
the same than the stringy
instanton contribution (\ref{ARGSP}).

\end{document}